\documentclass[twoside,leqno,twocolumn]{article}  
\usepackage{ltexpprt} 
\usepackage{amssymb}
\usepackage[fleqn]{amsmath}
\usepackage{bm}
\usepackage{xspace}
\usepackage{algorithm}
\usepackage{algpseudocode}
\usepackage{varwidth}
\usepackage{threeparttable}
\usepackage{multirow,bigdelim}
\usepackage{booktabs}
\usepackage{dsfont}
\usepackage{graphicx}
\usepackage{setspace}
\usepackage{hyperref}
\usepackage{xspace}
\usepackage{color}

\DeclareMathOperator{\sml}{sim}
\DeclareMathOperator{\gsml}{gsim}
\DeclareMathOperator{\imp}{imp}

\newcommand*{\CF}{\textsc{FBSM}\xspace}
\newcommand*{\CFNOSP}{\textsc{FBSM}}
\newcommand*{\CFEXP}{Feature-based factorized Bilinear Similarity Model\xspace}
\newcommand*{\CFEXPB}{\textbf{F}eature-based factorized \textbf{B}ilinear \textbf{S}imilarity \textbf{M}odel\xspace}

\newcommand*{\CFLIN}{\textsc{UFSM}\xspace}
\newcommand*{\CFLINNOSP}{\textsc{UFSM}}
\newcommand*{\CFLINEXP}{User-specific Feature-based Similarity Models\xspace}

\newcommand*{\TOPN}{Top-$n$\xspace}

\newcommand*{\RLFM}{\textsc{RLFM}\xspace}
\newcommand*{\RLFMI}{\textsc{RLFMI}\xspace}
\newcommand*{\RECN}{Rec@$n$\xspace}
\newcommand*{\DCGN}{DCG@$n$\xspace}
\newcommand*{\COSIM}{CoSim\xspace}

\newcommand*{\CUL}{CUL\xspace}
\newcommand*{\AMAZON}{AMAZON\xspace}
\newcommand*{\BX}{BX\xspace}
\newcommand*{\MLIMDB}{ML-IMDB\xspace}

\begin{document}

\title{\Large \CFEXP for Cold-Start \TOPN Item Recommendation\thanks{This work was
supported in part by NSF (IIS-0905220, OCI-1048018, CNS-1162405,
IIS-1247632, IIP-1414153, IIS-1447788), Army Research Office (W911NF-14-1-0316),
Samsung Research America, and the Digital Technology Center at the
University of Minnesota.}}
\author{Mohit Sharma~\thanks{University of Minnesota.}, Jiayu Zhou~\thanks{Samsung Research America}, Junling Hu$^\ddag$, George Karypis$^\dag$}
\date{}

\maketitle


\begin{abstract} \small\baselineskip=9pt 
Recommending new items to existing users has remained a challenging problem
due to absence of user's past preferences for these items. 
The user personalized non-collaborative methods based on item features 
can be used to address this item \emph{cold-start} problem. 
These methods rely on similarities between the target item and user's previous preferred items. 
While computing similarities based on item features, these
methods overlook the interactions among the features of the items and 
consider them independently. Modeling interactions among features can be helpful 
as some features, when considered together, provide a stronger signal on the 
relevance of an item when compared to case where features are considered independently. 
To address this important issue, in this work we introduce the \CFEXP (\CF), which learns factorized bilinear
similarity model for \TOPN recommendation of new items, given the
information about items preferred by users in past as well as the features of
these items. We carry out extensive empirical evaluations on benchmark
datasets, and we find that the proposed \CF approach improves upon
traditional non-collaborative methods in terms of recommendation performance.
Moreover, the proposed approach also learns insightful interactions among item
features from data, which lead to deep understanding on how these interactions contribute 
to personalized recommendation.
\end{abstract}

\section{Introduction}

\TOPN recommender systems are used to identify from a large pool of items those $n$
items that are the most relevant to a user and have become an essential
personalization and information filtering technology. They rely on the historical
preferences that were either explicitly or implicitly provided for the items and
typically employ various machine learning methods to build \emph{content-agnostic}
predictive models from these preferences. However, when new items are introduced into
the system, these approaches cannot be used to compute personalized recommendations,
because there are no prior preferences associated with those items.
As a result, the methods used to recommend new items, referred to as (item)
\emph{cold-start} recommender systems, in addition to the historical information,
take into account the characteristics of the items being recommended; that is, they
are \emph{content aware}. The items' characteristics are typically captured by a set
of domain-specific features. For example, a movie may have features like genre,
actors, and plot keywords; a book typically has features like content description and
author information. These item features are intrinsic to the item and as such they do
not depend on historical preferences. 

Over the years, a number of approaches have been developed towards solving the
item cold-start problem ~\cite{r1,r2,r10} that exploit the
features of the new items and the features of
the items on which a user has previously expressed his interest. A recently introduced approach, which
was shown to outperform other approaches is the \CFLINEXP (\CFLIN) \cite{r42}.
In this approach, a linear similarity function is estimated for each user that
depends entirely on features of the items previously liked by the user, which is
then used to compute a score indicating how relevant a new item will be for that
user. In order to leverage information across users (i.e., the transfer learning
component that is a key component of collaborative filtering), each user
specific similarity function is computed as a linear combination of a small
number of \emph{global} linear similarity functions that are shared across
users. Moreover, due to the way that it computes the preference scores, it can
achieve a high-degree of personalization while using only a very small number of
global linear similarity functions.

In this work we extend \CFLIN in order to account for interactions between the
different item features. We believe that such interactions are
important and quite common. For example, in an e-commerce website, the items
that users tend to buy are often designed to go well with previously purchased
items (e.g., a pair of shoes that goes well with a dress). The set of features
describing items of different type will be different (e.g., shoe material and
fabric color) and as such a linear model can not learn from the data that for
example a user prefers to wear leather shoes with black clothes. Being able to model such
dependencies can lead to item cold-start recommendation algorithms that achieve
better performance.

Towards this goal, we present a method called \CFEXPB (\CF) that uses bilinear
model to capture pairwise dependencies between the features.
Like \CFLIN, \CF learns a similarity function for estimating the similarity 
between items based on their features. However, unlike \CFLIN's linear global
similarity function, \CF's similarity function is bilinear.
A challenge associated with such bilinear models is that the number of
parameters that needs to be estimated becomes quadratic on the dimensionality of
the item's feature space, which is problematic given the very sparse training
data. \CF overcomes this challenge by assuming that the pairwise relations can
be modeled as a combination of a linear component and a low rank component. The
linear component allows it to capture the direct relations between the features
whereas the low rank component allows it to capture the pairwise
relations. 
The parameters of these models are
estimated using stochastic gradient descent and a ranking loss function based on
Bayesian Personalized Ranking (BPR) that optimizes the area under the receiver
operating characteristic curve.

We performed extensive empirical studies to evaluate the performance of the
proposed \CF on a variety benchmark datasets and compared it against state-of-
the-art models for cold-start recommendation, including latent factor methods
and non-collaborative user-personalized models. In our results \CF optimized
using BPR loss function outperformed other methods in terms of recommendation
quality.

\section{Notations and Definitions}
Throughout the paper, all vectors are column vectors and are represented by
bold lowercase letters (e.g., $\bm{f}_i$). Matrices are represented by upper
case letters (e.g., ${R,P,Q}$).

The historical preference information is represented by a preference matrix $R$.
Each row in ${R}$ corresponds to a user and each column corresponds to an item. 
The entries of $R$ are binary, reflecting user preferences on items. 
The preference given by user $u$ for item $i$ is represented by entry $r_{u,i}$ in $R$.  
The symbol $\tilde{r}_{u,i}$ represents the score predicted by the model for the actual
preference $r_{u,i}$.

Sets are represented with calligraphic letters. The set of users $\mathcal{U}$
has size $n_{\cal U}$, and the set of items $\mathcal{I}$ has a size $n_{\cal I}$.
$\mathcal{R}_u^+$ represents the set of items that user $u$ liked (i.e., $
\forall \, i \in \mathcal{R}_u^+, r_{u,i} = 1$). $\mathcal{R}_u^-$ represents
the set of items that user $u$ did not like or did not provide feedback for
(i.e., $ \forall \, i \in \mathcal{R}_u^-, r_{u,i} = 0$).

Each item has a feature vector that represents intrinsic characteristics
of that item. The feature vectors of all items are represented as the matrix
$F$ whose columns $\bm{f}_i$ correspond to the item feature vectors. The total
number of item features is $n_F$.

The objective of the \TOPN recommendation problem is to identify among the items
that the user has not previously seen, the $n$ items that he/she will like.

\section{Related Work}

The prior work to address the cold-start item recommendation can be divided 
into \emph{non-collaborative} user personalized models 
and \emph{collaborative} models. The non-collaborative models generate recommendations 
using only the user's past
interaction history and the collaborative models combine information from
the preferences of different users. 

Billsus and Pazzani~\cite{r23} developed one of the first user-modeling
approaches to identify relevant new items. In this approach they used the users'
past preferences to build user-specific models to classify new items
as either ``relevant'' or ``irrelevant''. The user models were built using
item features e.g., lexical word features for articles. Personalized 
user models~\cite{r24} were also used to classify news feeds by modeling short-term user needs 
using text-based features of items that were recently viewed by user and long-term needs were
modeled using news topics/categories. Banos~\cite{r26}
used topic taxonomies and synonyms to build high-accuracy 
content-based user models. 

Recently collaborative filtering techniques using latent
factor models have been used to address cold start item recommendation
problems. These techniques incorporate item features in their factorization
techniques. Regression-based latent factor models (RLFM)~\cite{r2} is a
general technique that can also work in item cold-start scenarios. 
RLFM learns a latent factor representation of the preference matrix
in which item features are transformed into a low dimensional space using regression.
This mapping can be used to obtain a low dimensional representation of the cold-start
items. User's preference on a new item is estimated by a dot product of
corresponding low dimensional representations. The RLFM model was further
improved by applying more flexible regression models~\cite{r10}.
AFM \cite{r1} learns item attributes to latent feature mapping
by learning a factorization of the preference matrix
into user and item latent factors $R = PQ^T$. A mapping function is then learned
to transform item attributes to a latent feature representation i.e., $R = PQ^T =
PAF^T$ where $F$ represents items' attributes and $A$ transforms the items' attributes 
to their latent feature representation. 


\CFLINEXP (\CFLIN)~\cite{r42} learns a personalized user
model by using historical preferences from all users across the dataset. In
this model for each user an item similarity function is learned, which is a
linear combination of user-independent similarity functions known as
\textit{global similarity functions}. Along with these global similarity
functions, for each user a personalized linear combination of these global
similarity functions is learned. It is shown to outperform both RLFM and AFM
methods in cold-start \TOPN item recommendations.

Predictive bilinear regression models \cite{r33} belong to the feature-based
machine learning approach to handle the cold-start scenario for both users and
items. Bilinear models can be derived from Tucker family \cite{r40}. They have
been applied to separate ``style'' and ``content'' in images\cite{r35}, to
match search queries and documents \cite{r37}, to perform semi-infinite stream
analysis~\cite{r41}, and etc. Bilinear regression models try to exploit the
correlation between user and item features by capturing the effect of pairwise
associations between them. Let $\bm{x}_i$ denotes features for user $i$ and $\bm{x}_j$
denotes features for item $j$, and a parametric bilinear indicator of the
interaction between them is given by $s_{ij} = \bm{x}_i^TW\bm{x}_j$ where $W$ denotes
the matrix that describes a linear projection from the user feature space onto
the item feature space. The method was developed for recommending cold-start
items in the real time scenario, where the item space is small but dynamic
with temporal characteristics. In another work~\cite{r38}, authors proposed to
use a pairwise loss function in the regression framework to learn the matrix
$W$, which can be applied to scenario where the item space is static but
large, and we need a ranked list of items. 

\section{Feature-based Similarity Model}
In this section we firstly introduce the feature-based linear model, analyzing 
the drawbacks of the model, and finally elaborate the technical details of our 
bilinear similarity model.

\subsection{Linear Similarity Models.} 

In \CFLIN \cite{r42} the preference score for new item $i$ for user $u$ is given by 
\begin{align*}
\tilde{r}_{u,i} 	& = \sum_{j\in \mathcal{R}_u^+} \sml_{u}(i,j), 
\end{align*}
\noindent where $\sml_u(i,j)$ is the user-specific similarity function given by
\begin{align*}
\sml_{u}(i,j)  & = \sum_{d=1}^{l} m_{u,d} \, \gsml_d(i,j), 
\end{align*}
\noindent where $\gsml_d(.)$ is the $d^{th}$ global similarity function, $l$ is the number
of global similarity functions,  and $m_{u,d}$ is a scalar that determines how
much the $d^{th}$ global similarity function contributes to $u$'s similarity
function. 

The similarity between two items $i$ and $j$ under the $d^{th}$ global similarity function $\gsml_d(.)$ is estimated as
\begin{align*}
\gsml_d(i,j) = \bm{w}_d (\bm{f}_i \odot \bm{f}_j)^T,
\end{align*}
where $\odot$ is the element-wise Hadamard product operator,  $\bm{f}_i$ and $\bm{f}_j$ are the feature vectors of items $i$ and $j$, respectively, and $\bm{w}_d$ is a vector of length $n_F$ with each entry $w_{d,c}$ holding the weight of feature $c$ under the global similarity function $\gsml_d(.)$. This weight reflects the contribution of feature $c$ in the item-item similarity estimated under $\gsml_d(.)$. Note that $\bm{w}_d$ is  a linear model on the feature vector resulting by the Hadamard product.

In author's  results \cite{r42} for datasets with large number of features only one
global similarity function was sufficient to outperform AFM and RLFM method for
\TOPN item cold-start recommendations. In
case of only one global similarity function the user-specific similarity
function is reduced to single global similarity function. Estimated preference
score for new item $i$ for user $u$ is given by 
\begin{align*}
    \tilde{r}_{u,i} 	& = \sum_{j\in \mathcal{R}_u^+} \sml_{u}(i,j) 
    = \sum_{j\in \mathcal{R}_u^+} \bm{w}_d (\bm{f}_i \odot \bm{f}_j)^T, 
\end{align*}
where $\bm{w}_d$ is the parameter vector, which can be estimated from training
data using different loss functions. 

\subsection{Factorized Bilinear Similarity Models.}\label{fbsm}

An advantage that the linear similarity method(\CFLIN) has, over state of art methods
such as RLFM and AFM, is that it uses information from all users across
dataset to estimate the parameter vector $\bm{w}_d$. As in the principle of
collaborative filtering, there exists users who have similar/dissimilar tastes
and thus being able to use information from other users can improve
recommendation for a user. However, we notice that a major drawback of this
model is that it fails to discover pattern affinities between 
item features. Capturing these correlations among features sometimes can lead
to significant improvements in estimating preference scores.

We thus propose \CF to overcome this drawback: It uses bilinear
formulation to capture correlation among item features. Similar to \CFLIN, it considers
information from all the users in dataset to learn these bilinear weights. In
\CF, the preference score for a new item $i$ for user $u$ is given by
\begin{equation} \label{est_Eq}
\begin{split}
\tilde{r}_{u,i} & = \sum_{j\in \mathcal{R}_u^+} \sml(i,j), \\
\end{split}
\end{equation}
\noindent where $\sml(i,j)$ is the similarity function given by
\begin{equation*}\label{simBiEq}
\begin{split}
  \sml(i,j)  & = \bm{f_i}^TW\bm{f_j} \\
\end{split}
\end{equation*}
\noindent where $W$ is the weight matrix which captures correlation among 
item features. Diagonal of matrix $W$ determines how well a feature of item
$i$ say $k^{th}$ feature of $i$ i.e., $f_{ik}$ interacts with corresponding
feature of item $j$ i.e., $f_{jk}$ while off-diagonal elements of $W$ gives
the contribution to similarity by interaction of item feature with other
features of item $j$ i.e., contribution of interaction between
$f_{ik}$ and $f_{jl}$ where $l \neq k$. Cosine similarity can be reduced to
our formulation where $W$ is a diagonal matrix with all the elements as ones.

A key challenge in estimating the bilinear model matrix $W$ is that the number
of parameters that needs to be estimated is quadratic in the number of features
used to describe the items. For low-dimensional datasets, this is not a
major limitation; however, for high-dimensional datasets, sometimes sufficient
training data is not present for reliable estimation.
This can become computationally infeasible, and moreover, lead to poor generalization
performance by overfitting the training data.
In order to overcome this drawback, we need to limit the degree of freedom of
the solution $W$, and we propose to represent $W$ as sum of diagonal weights
and low-rank approximation of the off-diagonal weights:
\begin{equation}
  \begin{split}
    W = D + V^TV
  \end{split}
\end{equation}
\noindent where $D$ is a diagonal matrix of dimension equal
to number of features whose diagonal is denoted using a vector $\bm{d}$, and
$V \in \mathbf{R}^{h \times n_F}$ is a matrix of rank $h$. The columns of $V$
represent latent space of features i.e., $\bm{v}_p$ represent latent factor of
feature $p$.
Using the low-rank approximation, the parameter matrix $W$ of the similarity
function is thus given by:
\begin{equation}\label{simEq}
\begin{split}
  \sml(i,j)  & = \bm{f}_i^TW\bm{f}_j = \bm{f}_i^T(D + V^TV)\bm{f}_j\\
   &= \bm{d} (\bm{f}_i \odot \bm{f}_j)^T + \sum_{k=1}^{n_F}\sum_{\substack{p=1 }}^{n_F} f_{ik}f_{jp}v_k^Tv_p\\
\end{split}
\end{equation}
The second part of equation \ref{simEq} captures the effect of off-diagonal
elements of $W$ by inner product of latent factor of features. Since we are
now estimating only diagonal weights and low-rank approximation of off-
diagonal weights, the computation reduces significantly compared to when we
were trying to estimate the complete matrix $W$. This also gives us a flexible
model where we can regularize diagonal weights and feature latent factors
separately.

The bilinear model may look similar to the formulation described
in~\cite{r33}, and however the two are very different in nature: in~\cite{r33}
the bilinear model is used to capture correlation among user and item
features, on contrary the \CF is trying to find correlation within features of
items itself. The advantage of modeling interactions among item features is
especially attractive when there is no explicit user features available. Note
that it is not hard to encode the user features in the proposed bilinear model
such that the similarity function is parameterized by user features, and we
leave a detailed study to an extension of this paper.

\subsection{Parameter Estimation of \CF. }

\CF is parameterized by $\Theta=[D, V]$, where
$D, V$ are the parameters of the similarity function. The inputs to the
learning process are: ({\romannumeral 1}) the preference matrix ${R}$,
({\romannumeral 2}) the item-feature matrix ${F}$, and ({\romannumeral 3}) the
dimension of latent factor of features. There are many loss functions we can
choose to estimate $\Theta$, among which the Bayesian Personalized Ranking
(BPR) loss function \cite{r6} is designed especially for ranking problems.
In the \TOPN recommender systems, the predicted preference scores are used
to rank the items in order to select the highest scoring $n$ items, and thus
the BPR loss function can better model the problem than other loss functions
such as least squares loss and in general lead to better empirical
performance~\cite{r1,r6}. As such, in this paper, we propose to use the BPR
loss function, and in this section we show how the loss function can be used
to estimate the parameters $\Theta$. Note that other loss functions such as
least squared loss can be applied similarity.

We denote the problem of solving \CF using BPR as FBSM$_{bpr}$, and the loss 
function is given by
\begin{equation}\label{eq_bpr}
\mathcal{L}_{bpr}(\Theta) \equiv - \sum_{u \in U} \sum_{\substack{i \in \mathcal{R}_u^+ ,\\  j \in \mathcal{R}_u^-}}  \ln \, \sigma(\tilde{r}_{u,i}(\Theta) - \tilde{r}_{u,j}(\Theta) ),
\end{equation}
where, $\tilde{r}_{u,i}$ is the predicted value of the user $u$'s preference for the item $i$
and $\sigma$ is the sigmoid function. 
The BPR loss function tries to learn item preference scores such that the items that a user likes have higher preference scores than the ones he/she does not like, regardless of the actual item preference scores. 
The prediction value $\tilde{r}_{u,i}$ is given by:
\begin{equation} \label{eq_est_true}
\tilde{r}_{u,i} = \sum_{ j \in \mathcal{R}_u^+ \backslash i} \sml_{u}(i,j),
\end{equation}
which is identical to Equation \ref{est_Eq} except that item $i$ is excluded from the summation. This is done to ensure that the variable being estimated (the dependent variable) is not used during the estimation as an independent variable as well. We refer to this as the \emph{Estimation Constraint}~\cite{r32}. 


To this end, the model parameters $\Theta=[D, V ]$ are estimated via an optimization process of the form:
\begin{equation} \label{eq_opt}
\begin{split}
  \min_{\Theta=[D, V ]} \mathcal{L}_{bpr}(\Theta) + \lambda \|V\|_F^2 + \beta\|D\|_F^2, \\
\end{split}
\end{equation}
where we penalize the frobenius norm of the model parameters in order to
control the model complexity and improve its generalizability.

To optimized Eq.~(\ref{eq_opt}) we proposed to use stochastic gradient
descent(SGD)~\cite{r22}, in order to handle large-scale datasets. The update
steps for $D, V$ are based on triplets $(u,i,j)$ sampled from training
data. For each triplet, we need to compute the corresponding estimated
relative rank $\tilde{r}_{u,ij} = \tilde{r}_{u,i} - \tilde{r}_{u,j}$. Let
\begin{align*}
\tau_{u,ij} = \text{sigmoid}(-\tilde{r}_{u,ij}) = \frac{e^{-\tilde{r}_{u,ij}} }  {1 + e^{- \tilde{r}_{u,ij}} },
\end{align*}
the updates are then given by:
\begin{equation} \label{dUpdBPREq}
\begin{split}
  D &=  D + \alpha_1 \Big( \tau_{u,ij}  \nabla_{D} \tilde{r}_{u,ij} -
 2 \beta D  \Big), and
\end{split}
\end{equation}
%
%
\begin{equation} \label{vUpdBPREq}
\begin{split}
  \bm{v}_{p} &=  \bm{v}_{p} + \alpha_2 \Big(  \tau_{u,ij} \nabla_{\bm{v}_{p}} \tilde{r}_{u,ij} -
 2 \lambda \bm{v}_p  \Big). 
\end{split}
\end{equation}
%
\subsection {Performance optimizations}
In our approach, the direct computation of gradients is
time-consuming and is prohibitive when we have high-dimensional item
features. For example, the relative rank $\tilde{r}_{u,ij}$ given by
\begin{equation}\label{relRankExpEq}
\begin{split}
\tilde{r}_{u,ij} &= \Bigg(\bm{f}_i^T (D + V^TV) \Big(\Big(\sum_{\substack{ q \in
\mathcal{R}_u^+ \backslash i}}\bm{f}_q\Big) - \bm{f}_i\Big) \Bigg) -  \\
&\quad  \Bigg( \bm{f}_j^T (D + V^TV)
\Big(\sum_{\substack{ q \in \mathcal{R}_u^+}}\bm{f}_q\Big) \Bigg) ,
\end{split}
\end{equation}
has complexity of $O(|\mathcal{R}_u^+|n_F h )$, where $h$ is the dimensionality
of latent factors, $n_F$ is the number of features.

To efficiently compute these, let 
\begin{equation*}
  \bm{f_u} = \sum_{\substack{ q \in \mathcal{R}_u^+}}\bm{f}_q, 
\end{equation*}
which can be precomputed once for all users.

Then, Equation \ref{relRankExpEq} becomes
\begin{equation*}
\begin{split}
  \tilde{r}_{u,ij} &= \Big( \bm{f}_i^T (D + V^TV) (\bm{f}_u - \bm{f}_i) \Big) -\\
                   &\quad \quad\Big( \bm{f}_j^T (D + V^TV) \bm{f}_u \Big)  \\
  &= \Big( (\bm{f}_i-\bm{f}_j)^TD\bm{f}_u - \bm{f}_i^TD\bm{f}_i \Big) +  \\
  &\quad \quad  \Big( (\bm{f}_i-\bm{f}_j)^T(V^TV)\bm{f}_u -
  \bm{f}_i^TV^TV\bm{f}_i\Big)  \\
  &= \Big( \bm{\delta}_{ij}^TD\bm{f}_u - \bm{f}_i^TD\bm{f}_i \Big) +\\
  &\quad \quad \Big(\bm{\delta}_{ij}^T(V^TV)\bm{f}_u -
  \bm{f}_i^TV^TV\bm{f}_i\Big)  \\
  &=  \Big( \bm{\delta}_{ij}^TD\bm{f}_u - \bm{f}_i^TD\bm{f}_i \Big) +\\
  &\quad \quad \Big((V\bm{\delta}_{ij})^T(V\bm{f}_u) -
  (V\bm{f}_i)^T(V\bm{f}_i)\Big)  ,
\end{split}
\end{equation*}

where $\bm{\delta}_{ij} = \bm{f}_i-\bm{f}_j$.

The complexity of computing the relative rank then becomes $O(n_F h)$, which
is lower than complexity of Equation \ref{relRankExpEq}.

The gradient of the diagonal component is given by
\begin{equation} \label{eqDiagGrad}
  \begin{split}
    \frac{\partial \tilde{r}_{u,ij}}{\partial D}  
    &= \Big(\bm{\delta}_{ij} \otimes \bm{f}_u - \bm{f}_i \otimes
  \bm{f}_i \Big).\\
  \end{split}
\end{equation}
where $\otimes$ represents elementwise scalar product. The complexity of
Equation \ref{eqDiagGrad} is given by $O(n_F)$.

The gradient of the low rank component is given by
\begin{equation}
  \begin{split}
    \frac{\partial \tilde{r}_{u,ij}}{\partial \bm{v}_p}  
    &= \bm{\delta}_{ij,p} (V f_u) +  \bm{f}_{up} (V \bm{\delta}_{ij}) - 2\bm{f}_{ip} (V f_i),\\ 
  \end{split}
\end{equation}
whose complexity is $O(n_F h)$. 

Hence, the complexity of gradient
computation for all the parameters is given by $O(n_F h + n_F )
\approx O(n_F h)$. We were able to obtain both the estimated relative rank
and all the gradients in $O(n_F h)$, which is linear with respect to
feature dimensionality as well as the size of latent factors and the number of
global similarity functions. This allows the $\CF_{bpr}$ to process large-scale datasets.


\begin{algorithm}[ht]
\caption{FBSM$_{bpr}$-Learn}
\label{alg-bpr}
\begin{algorithmic}[1]
  \Procedure{\CF$_{bpr}$\_Learn}{}
    \State $\lambda \gets$ V regularization weight
    \State $\beta \gets$ D regularization weight
    \State $\alpha_1, \alpha_2 \gets$ D and V learning rates
    \State Initialize $\Theta=[D, V]$ randomly 
    
    \State
    
    \While {not converged}
      \For{ each user $u$} 
        \State sample a pair ($i,j$) s.t.  $i \in \mathcal{R}_u^+$, $j \in \mathcal{R}_u^-$
        \State compute $\tilde{r}_{u,ij} = \tilde{r}_{u,i} - \tilde{r}_{u,j}$
        \State compute $\nabla_{D} \tilde{r}_{u,ij} $ 
        \State compute $\nabla_{\bm{v}_{p}} \tilde{r}_{u,ij} $ 
        
        \State update $D$ using (\ref{dUpdBPREq})
        \State update $\bm{v}_p \forall p$  using (\ref{vUpdBPREq})
      \EndFor
    \EndWhile
    \State
    \State \Return $\Theta=[D,V]$
  \EndProcedure
\end{algorithmic}
\end{algorithm}

We note that the proposed FBSM$_{bpr}$ method is closely related to the
factorization machine (FM)~\cite{r3}, in that both are exploring the
interactions among the features.  However, there is one key difference between
these two: while the FM is heavily dependent on the quality of the user
features, the proposed method does not depend on such user features.

\section{Experimental Evaluation}

In this section we perform experiments to 
demonstrate the effectiveness of the proposed algorithm. 

\subsection{Datasets}
We used four datasets (Amazon Books, MovieLens-IMDB, CiteULike, Book Crossing) to evaluate the performance of \CF.

Amazon Books(\AMAZON) is a dataset collected from amazon about best-selling books
and and their ratings. The ratings are binarized by treating all ratings greater
than equal to $3$ as $1$ and ratings below $3$ as $0$. Each is accompanied with a
description which was used as item's content.

CiteULike(\CUL)\footnote{http://citeulike.org/} aids researchers by allowing
them to add scientific articles to their libraries. For users of the \CUL, the
articles present in their library are considered as preferred articles i.e. $1$
in a preference matrix while rest are considered as implicit $0$ preferences. 

MovieLens-IMDB (\MLIMDB) is a dataset extracted from the IMDB and the
MovieLens-1M datasets\footnote{http://www.movielens.org, http://www.imdb.com} by
mapping the MovieLens and IMDB movie IDs and collecting the movies that have
plots and keywords. The ratings were binarized similar to \AMAZON by treating
all ratings greater than $5$ as $1$ and below or equal to $5$ as 0. The movies 
plots and keywords were used as the item's content. 
Book Crossing (\BX) dataset is extracted from Book Crossing data \cite{r31} such
that user has given at least four ratings and each book has received the same
amount of ratings. Description of these books were collected from Amazon using
ISBN and were used as item features.

For the \AMAZON, \CUL, \BX and \MLIMDB datasets, the words that appear in the item descriptions were collected, stop words were removed and the remaining words were stemmed to generate the terms that were used as the item features. All words that appear in less than 20 items and all words that appear in more than 20\% of the items were omitted. The remaining words were represented with TF-IDF scores. 

Various statistics about these datasets are shown in Table \ref{datasets_table}. Most of these datasets contain items that have high-dimensional feature spaces. Also comparing the densities of the datasets we can see that the MovieLens dataset have significantly higher density than other dataset.


\begin{table*}[thb]
   \centering
  \label{datasets_table}
    \caption{Statistics for the datasets used for testing}  
    
   
    \begin{threeparttable}
    \begin{tabular}{
		@{\hspace{2pt}}l@{\hspace{3pt}}
		@{\hspace{2pt}}r@{\hspace{3pt}}
		@{\hspace{2pt}}r@{\hspace{3pt}}
		@{\hspace{2pt}}r@{\hspace{3pt}}
		@{\hspace{2pt}}r@{\hspace{3pt}}
		@{\hspace{2pt}}r@{\hspace{3pt}}
		@{\hspace{2pt}}r@{\hspace{3pt}}
		@{\hspace{2pt}}r@{\hspace{3pt}}
	}
	\hline
	\hline
	\textbf{Dataset}	&	\textbf{\# users}	&	\textbf{\# items}		&	\textbf{\# features}	&	\textbf{\# preferences}	&	\textbf{\# prefs/user}	&	\textbf{\# prefs/item}	& \textbf{density}\\
	\hline
	\hline
	\rule{-2pt}{3ex}
	\CUL  		& 	3,272 & 	21,508 	& 	6,359 	& 	180,622 	&	55.2 		& 	8.4 		&  	0.13\%	\\
	\rule{-2pt}{3ex}
	\BX & 17,219 &	36,546 	& 	8,946 	& 	574,127 	& 	33.3 		& 	15.7 		&	0.09\%	\\
	\rule{-2pt}{3ex}
  \AMAZON		 & 13,097 	& 	11,077 	& 	5,766 	& 	175,612 	& 	13.4 		& 	15.9	&	0.12\%	\\ 
	\rule{-2pt}{3ex}
 	\MLIMDB 		&	2,113& 	8,645 	& 	8,744 	& 	739,973 	& 	350.2 	& 	85.6 		&	4.05\%	\\
	\hline
	\hline
	
	\end{tabular}

    \end{threeparttable}
\end{table*}


\subsection{Comparison methods} \label{other_methods}
We compared \CF against non-collaborative personalized user modeling methods and
collaborative methods.

\begin{enumerate}
  \item{\textbf{Non-Collaborative Personalized User Modeling Methods}}
    Following method is quite similar to method described in \cite{r23}
    \begin{itemize}
      \item \textbf{Cosine-Similarity (\COSIM):} This is a personalized
        user-modeling method.
        The preference score of user $u$ on target item $i$ is estimated using
        equation \ref{eq_est_true} by using \emph{cosine similarity} between
        item features.
    \end{itemize}
  \item{\textbf{Collaborative Methods}}
    \begin{itemize}
      \item \textbf{\CFLINEXP (\CFLIN):}
        As mentioned before, this method~\cite{r42} learns personalized user model
        by using all past preferences from users across the dataset. It
        outperformed other state of the art collaborative latent factor based methods e.g.,
        RLFM\cite{r2},  AFM\cite{r1} by significant margin.
      \item{\textbf{\RLFMI}:} 
        We used the Regression-based Latent Factor Modeling(\RLFM) technique
        implemented in factorization machine library LibFM\cite{r3} that 
        accounts for inter-feature interactions. We used LibFM with SGD learning
        to obtain \RLFMI results. 
    \end{itemize}
\end{enumerate}

\begin{table*}[t!]
  \caption {Performance of \CF and Other Techniques on the Different Datasets } \label{table1}    
  \begin{center}
    \vspace{-1pt}
   
    \begin{threeparttable}
    \begin{tabular} {
      @{\hspace{0pt}}l@{\hspace{1pt}}
      @{\hspace{1pt}}p{1.7cm}@{\hspace{2pt}}
      @{\hspace{1pt}}r@{\hspace{2pt}}
      @{\hspace{1pt}}r@{\hspace{2pt}}
      @{\hspace{1pt}}r@{\hspace{2pt}}
      @{\hspace{1pt}}p{1.757cm}@{\hspace{1pt}}
      @{\hspace{1pt}}r@{\hspace{2pt}}
      @{\hspace{1pt}}r@{\hspace{2pt}}
      @{\hspace{1pt}}r@{\hspace{2pt}}
      @{\hspace{1pt}}r@{\hspace{2pt}}
      @{\hspace{1pt}}r@{\hspace{1pt}}
      }
  \hline
  \hline

  \multirow{2}{*}{\bf{Method}}  & \multicolumn{3}{c}{\bf{\CUL}} &
              & \multicolumn{3}{c}{\bf{\BX}} \\
  \cmidrule(r){2-4} 
  \cmidrule(r){6-8}
  & Params  & Rec@10  & DCG@10    &
  & Params  & Rec@10  & DCG@10    \\
  \hline
  \rule{-2pt}{3ex} 
  \COSIM      & -   & 0.1791  & 0.0684   &
          & -   & 0.0681  & 0.0119  \\
  \hline
  \rule{-2pt}{3ex} 
  \RLFMI  & h=75 & 0.0874 & 0.0424 &
          & h=75 & 0.0111 & 0.003 & \\
  \hline
  \rule{-2pt}{3ex}
  \CFLINNOSP$_{bpr}$   & $l$=1, $\mu_1$=0.25 & 0.2017 & 0.0791   &
          & $l$=1, $\mu_1$=0.1  & 0.0774  & 0.0148 & \\
  \hline
  \rule{-2pt}{3ex}
  \CFNOSP$_{bpr}$    & $\lambda$=0.25, $\beta$=10, h=5   & \underline{0.2026}  & \underline{0.0792}  &
          & $\lambda$=1, $\beta$=100, h=1 & \underline{0.0776} & \underline{0.0148} & \\ 
  \hline
  
  \end{tabular}
  \begin{tabular} {
      @{\hspace{0pt}}l@{\hspace{1pt}}
      @{\hspace{1pt}}p{1.7cm}@{\hspace{2pt}}
      @{\hspace{1pt}}r@{\hspace{2pt}}
      @{\hspace{1pt}}r@{\hspace{2pt}}
      @{\hspace{1pt}}r@{\hspace{2pt}}
      @{\hspace{1pt}}p{1.757cm}@{\hspace{1pt}}
      @{\hspace{1pt}}r@{\hspace{2pt}}
      @{\hspace{1pt}}r@{\hspace{2pt}}
      @{\hspace{1pt}}r@{\hspace{2pt}}
      @{\hspace{1pt}}r@{\hspace{2pt}}
      @{\hspace{1pt}}r@{\hspace{1pt}}
      }

  \hline
  \hline

  \multirow{2}{*}{\bf{Method}} &  \multicolumn{3}{c}{\bf{\MLIMDB}} & \multicolumn{3}{c}{\bf{\AMAZON}}\\
  \cmidrule(r){2-4} 
  \cmidrule(r){6-8}
  & Params  & Rec@10  & DCG@10 & 
  & Params  & Rec@10 &  DCG@10  \\
  \hline
  \rule{-2pt}{3ex} 
  \COSIM  & -   & 0.0525  & 0.1282 &
          & -   & 0.1205  & 0.0228 & \\
          \hline
          \rule{-2pt}{3ex}
          \RLFMI  & h = 15 & 0.0155 & 0.0455 & 
          & h = 30 & 0.0394 & 0.0076 \\
          \hline
          \rule{-2pt}{3ex}
          \CFLINNOSP$_{bpr}$ & $l$=1, $\mu_1$=0.005  & 0.0937 & 0.216 &
            & $l$=1, $\mu_1$=0.25, & 0.1376  & 0.0282  & \\ 
          \hline
          \rule{-2pt}{3ex}
  \CFNOSP$_{bpr}$ &  $\lambda$=0.01, $\beta$=0.1, h=5  & \underline{0.0964}  &
\underline{0.227} &
            & $\lambda$=0.1, $\beta$=1, h=1 & \underline{0.1392}  & \underline{0.0284}  & \\ 
  \hline
  
  \end{tabular}
  
  \begin{tablenotes}
    \item[]\scriptsize
  The ``Params'' column shows the main parameters for each method.  
  For \CFLIN$_{bpr}$, $l$ is the number of similarity functions, and $\lambda$,
  $\mu_1$ is the regularization parameter.
  For $\CF, \lambda$ and $\beta$ are regularization parameters and h is
  dimension of feature latent factors. 
  The ``Rec@10'' and ``DCG@10'' columns show the values obtained for these evaluation metrics. 
  The entries that are underlined represent the best performance obtained for each dataset. 
    \end{tablenotes}

\end{threeparttable}
\end{center}
\end{table*}

\subsection{Evaluation Methodology and Metrics}
We evaluated performance of methods using the following procedure. 
For each dataset we split the corresponding user-item preference matrix  $R$ into three
matrices $R_{train}$, $R_{val}$ and $R_{test}$. $R_{train}$ contains a randomly
selected $60\%$ of the columns (items) of R, and the remaining columns were divided equally
among $R_{val}$ and $R_{test}$. Since items in $R_{test}$ and $R_{val}$ are not
present in $R_{train}$, this allows us to evaluate the methods for item cold-start
problems as users in $R_{train}$ do not have any preferences for items in
$R_{test}$ or $R_{val}$. 
The models are learned using $R_{train}$ and the best model is
selected based on its performance on the validation set $R_{val}$. The selected model
is then used to estimate the preferences over all items in $R_{test}$. For each
user the items are sorted in decreasing order and the first $n$ items are returned
as the \TOPN recommendations for each user. The evaluation metrics as described
later are computed using these \TOPN recommendation for each user. 

After creating the train, validation and test split, there might be some users who do
not have any items in validation or test split. In that case we evaluate performance on
the splits for only those user who have at least one item in corresponding test split.
This split-train-evaluate procedure is repeated three times for each dataset and
evaluation metric scores are averaged over three runs before being reported in
results.

We used two metrics to assess the performance of the various methods: Recall at
$n$ (Rec@n) and Discounted Cumulative Gain at $n$ (DCG@n).
Given the list of the \TOPN recommended items for user $u$, \RECN measures how
many of the items liked by $u$ appeared in that list, whereas the \DCGN measures
how high the relevant items were placed in the list. 
The \RECN is defined as 
\begin{equation*}
  REC@n = \frac{|\{\mbox{Items liked by user}\} \cap \{\mbox{Top-n items}\}|}{|\mbox{Top-n items}|}
\end{equation*}

The \DCGN is defined as
\begin{equation*}
  DCG@n= \imp_1 + \sum_{p=2}^{n} \frac{\imp_p}{\log_2(p)},
\end{equation*} 
where the importance score $\imp_p$ of the item with rank $p$ in the \TOPN list is
\begin{equation*}
\imp_p = \left\{
  \begin{array}{l l}
    1/n, & \quad \text{if item at rank $p \in R_{u,test}^+$}\\
    0, & \quad \text{if item at rank $p \notin R_{u,test}^+$}.
  \end{array} \right. 
  \end{equation*} 
The main difference between \RECN and \DCGN is that \DCGN is sensitive to the rank of the 
items in the \TOPN list. Both the \RECN and the \DCGN are computed for each user and then 
averaged over all the users.

\subsection{Model Training} \label{evaluation}
\CF's model parameters are estimated using training set $R_{train}$ and
validation set $R_{val}$. After each
major SGD iteration of Algorithm \ref{alg-bpr} we compute
the \RECN on validation set and save the current model if current \RECN is 
better than those computed in previous iterations. The learning process ends when
the optimization objective converges or no further improvement in validation recall 
is observed for $10$ major SGD iterations. At the end of learning process we return 
the model that achieved the best \RECN on the validation set.

To estimate the model parameters of $\CF_{bpr}$, we draw samples
equal to the number of preferences in $R$ for each major SGD iteration. Each sample
consists of a user, an item preferred by user and an item not preferred by user.
If a dataset does not contain items not preferred by user then we sample from
items for which his preference is not known.

For estimating \RLFMI model parameters, LibFM was given the training and
validation sets and the model that performed best on the validation set 
was used for evaluation on test sets. For \RLFMI, the training set must 
contain both 0's and 1's. Since the
\CUL dataset does not contain both 0's and 1's, we sampled 0's equal to number
of 1's in $R$ from the unknown values.

\section{Results and Discussion}
\subsection{Comparison with previous methods}

We compared the performance of \CF with other methods described in Section on
\ref{other_methods}.
Results are shown in Table \ref{table1} for different datasets. We tried different
values for various parameters e.g., latent factors and regularization parameters
associated with methods and report the best results found across datasets.

These results illustrate that $\CF_{bpr}$ by modeling the
cross feature interactions among items can improve upon the $\CFLIN$
method \cite{r42} which has been shown to outperform the existing 
state of the art methods like RLFM\cite{r2} and AFM\cite{r1}. Similar to the
$\CFLIN$ method, $\CF_{bpr}$ method has outperformed latent-factor based \RLFMI
method. An example of cross-feature interactions
found by \CF is interaction among terms \textit{tragic, blockbuster,
}and \textit{famous}.

\subsection{Performance investigation at user level}
We further looked at some of our datasets ($\MLIMDB$ and $\AMAZON$) and divided the users
based on the performance achieved by $\CF$ in comparison with $\CFLIN$ i.e.,
users for which $\CF$ performed better, similar and worse than $\CFLIN$. These
finding are presented in Table \ref{USER_INV}. For \MLIMDB
dataset there is an increase of $22\%$ in number of users for whom recommendation
is better on using $\CF$ method, while for $\AMAZON$ dataset the number of users
that benefited from $\CF$ is not significant. On comparing the two datasets in
Table \ref{USER_INV}, $\MLIMDB$ has much more preferences per item
or existing items have been rated by more users compared to $\AMAZON$. 
Hence our proposed method $\CF$ takes the advantage of availability of more data while 
$\CFLIN$ fails to do so.

\begin{table*}[tb]
  \centering
  \caption{User level investigation for \MLIMDB and \AMAZON}\label{USER_INV}
  \begin{tabular}{ p{2cm} p{2cm} c c p{2.0cm} p{2.0cm} }
    \hline
    \textbf{Dataset} &\textbf{$\CF$ against $\CFLIN$} & \textbf{users} &
      \textbf{items} & \textbf{average user preferences} & \textbf{average item
  preferences} \\ 
    \hline
    \multirow{3}{*}{\MLIMDB} & BETTER & 887 & 4770 & 224 & 42 \\ 
                             & SAME & 802 & 4371 & 119 & 22 \\ 
                             & WORSE & 424 & 3928 & 137 & 15 \\ 
    \hline
    \multirow{3}{*}{\AMAZON} & BETTER &  325 &  4170 &  23& 2 \\ 
                             & SAME & 12458 & 6638 & 7 & 13 \\ 
                             & WORSE & 314 & 4294 & 24 & 2 \\ 
    \hline
  \end{tabular}
\end{table*}

\section{Conclusion}
We presented here FBSM for \TOPN recommendation in item cold-start scenario.
It tries to learn a similarity function between items, represented by their features, by
using all the information available across users and also tries to capture
interaction between features by using a bilinear model.
Computation complexity of bilinear model estimation is significantly reduced by
modeling the similarity as sum of diagonal component and off-diagonal component.
Off-diagonal component are further estimated as dot product of latent spaces of
features.


In future, we want to investigate the effect of non-negativity constraint on
model parameters and effectiveness of the method on actual rating prediction
instead of \TOPN recommendation.

\bibliography{refs}{}
\bibliographystyle{plain}

\clearpage
\newpage
\onecolumn

\end{document}